\documentclass[sigconf]{acmart}

\usepackage{subcaption}
\AtBeginDocument{%
  \providecommand\BibTeX{{%
    \normalfont B\kern-0.5em{\scshape i\kern-0.25em b}\kern-0.8em\TeX}}}

\copyrightyear{2023} 
\acmYear{2023} 
\setcopyright{acmlicensed}\acmConference[SIGIR '23]{Proceedings of the 46th International ACM SIGIR Conference on Research and Development in Information Retrieval}{July 23--27, 2023}{Taipei, Taiwan}
\acmBooktitle{Proceedings of the 46th International ACM SIGIR Conference on Research and Development in Information Retrieval (SIGIR '23), July 23--27, 2023, Taipei, Taiwan}
\acmPrice{15.00}
\acmDOI{10.1145/3539618.3591852}
\acmISBN{978-1-4503-9408-6/23/07}

\begin{document}

\title{Extracting Complex Named Entities in Legal Documents via Weakly Supervised Object Detection}

\author{Hsiu-Wei Yang}
\authornote{Both authors contributed equally to this research.}
\affiliation{%
  \institution{Thomson Reuters Labs}
  \streetaddress{333 Bay St.}
  \state{Ontario}
  \city{Toronto}
  \country{Canada}
  \postcode{M5H 2R2}
}
\email{leo.yang@thomsonreuters.com}

\author{Abhinav Agrawal}
\authornotemark[1]
\affiliation{%
  \institution{Thomson Reuters Labs}
  \city{Bangalore}
  \state{Karnataka}
  \country{India}
}
\email{abhinav.agrawal@thomsonreuters.com}

\renewcommand{\shortauthors}{Yang and Agrawal}

\begin{abstract}
Accurate Named Entity Recognition (NER) is crucial for various information retrieval tasks in industry. However, despite significant progress in traditional NER methods, the extraction of Complex Named Entities remains a relatively unexplored area. In this paper, we propose a novel system that combines object detection for Document Layout Analysis (DLA) with weakly supervised learning to address the challenge of extracting discontinuous complex named entities in legal documents. Notably, to the best of our knowledge, this is the first work to apply weak supervision to DLA. Our experimental results show that the model trained solely on pseudo labels outperforms the supervised baseline when gold-standard data is limited, highlighting the effectiveness of our proposed approach in reducing the dependency on annotated data.
\end{abstract}






\maketitle

\section{Introduction}
Named Entity Recognition (NER) is a long-studied field in Natural Language Processing (NLP) that involves detecting and classifying named entity (NE) mentions. 
By correctly identifying NEs in queries or documents, search engines can improve both relevancy and efficiency \cite{voorhees2005trec, guo2009named, khalid2008impact}. 
For example, in the legal domain, knowing which paragraphs contain attorney information can help narrow the search scope for attorney-related queries. 
However, traditional NER techniques may not be suitable for the legal NER problems. 
In the industry, it is common to see the NEs of interest fall under the challenging category of \textit{Complex Named Entity}, 
which is described as \textit{``nested, overlapping, and discontinuous''} \cite{dai2018recognizing} or \textit{``semantically ambiguious''} \cite{malmasi2022semeval}.

\begin{figure}[ht!]
    \centering
    \begin{subfigure}[h]{0.42\textwidth}
        \centering
        \includegraphics[width=\linewidth]{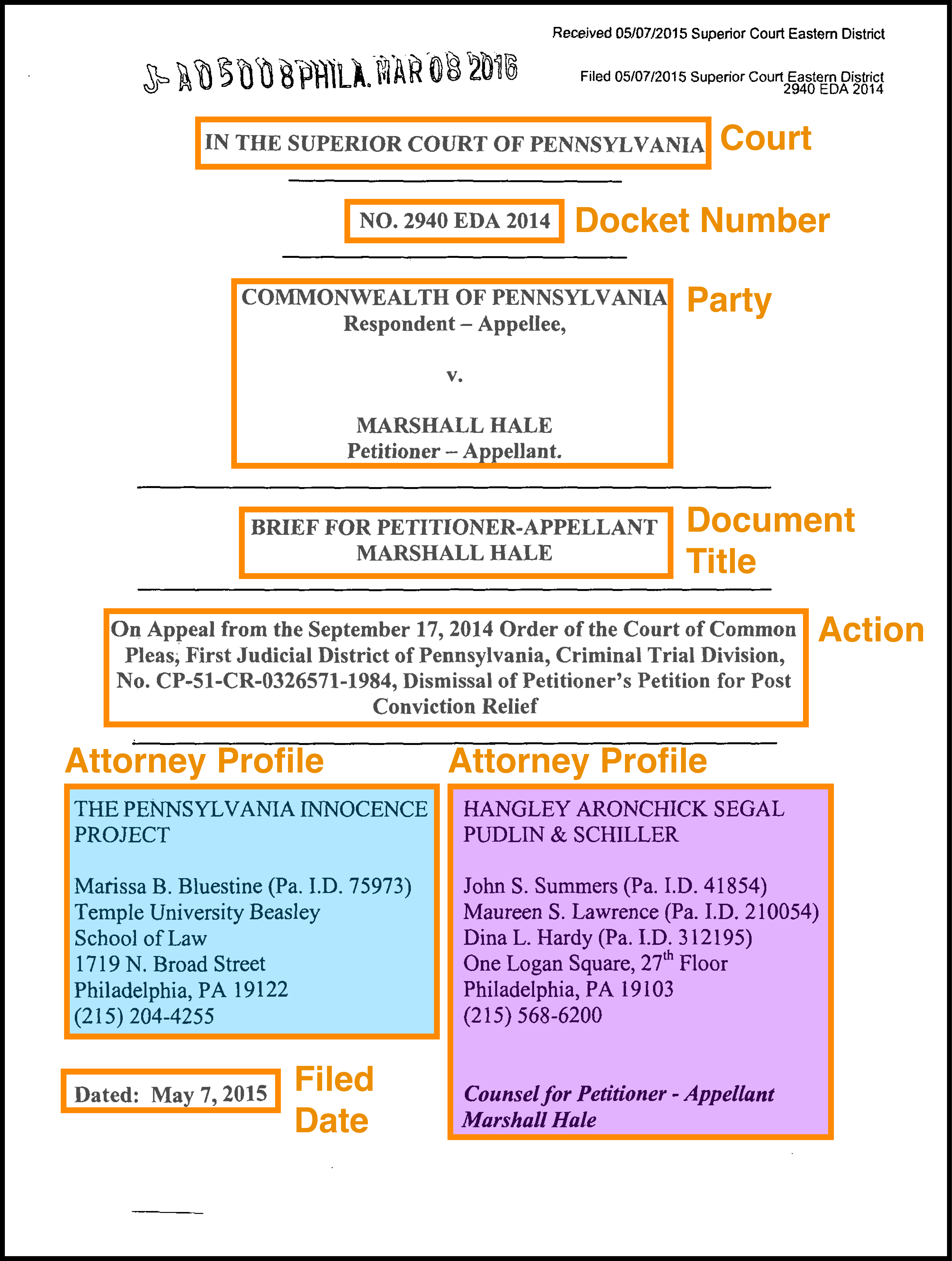}
        \caption{Caption of the Brief}
        \label{fig:brief}
    \end{subfigure} 
    \hfill
    \begin{subfigure}[h]{0.42\textwidth}
        \centering
        \includegraphics[width=\linewidth]{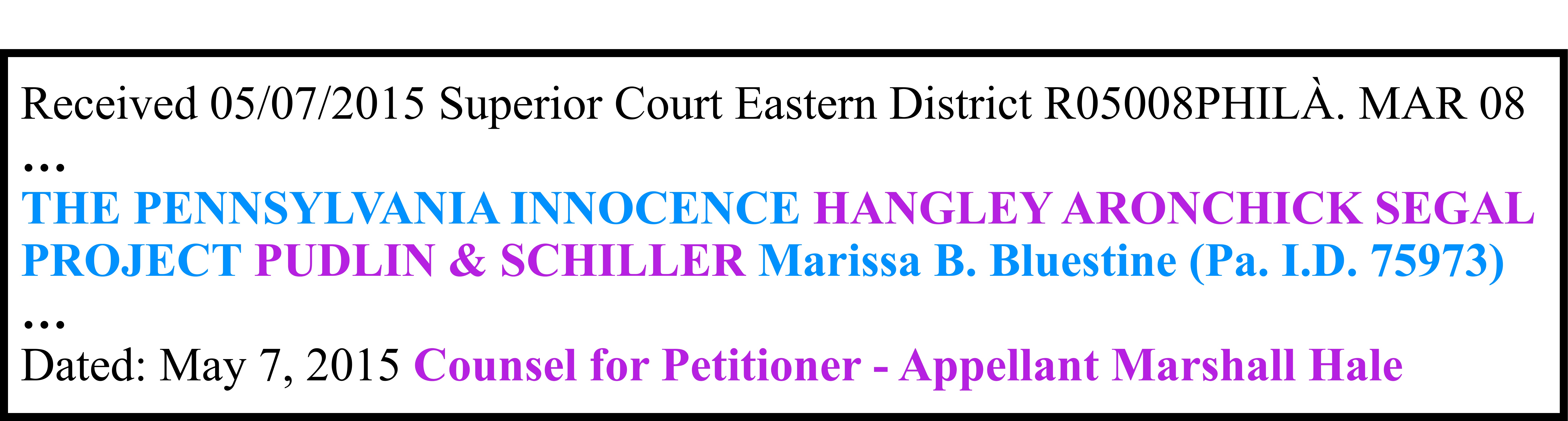}
        \caption{Complexity in OCR'd Text}
        \label{fig:ocr}
    \end{subfigure}
    \hfill
    \begin{subfigure}[h]{0.34\textwidth}
        \centering
        \includegraphics[width=\linewidth]{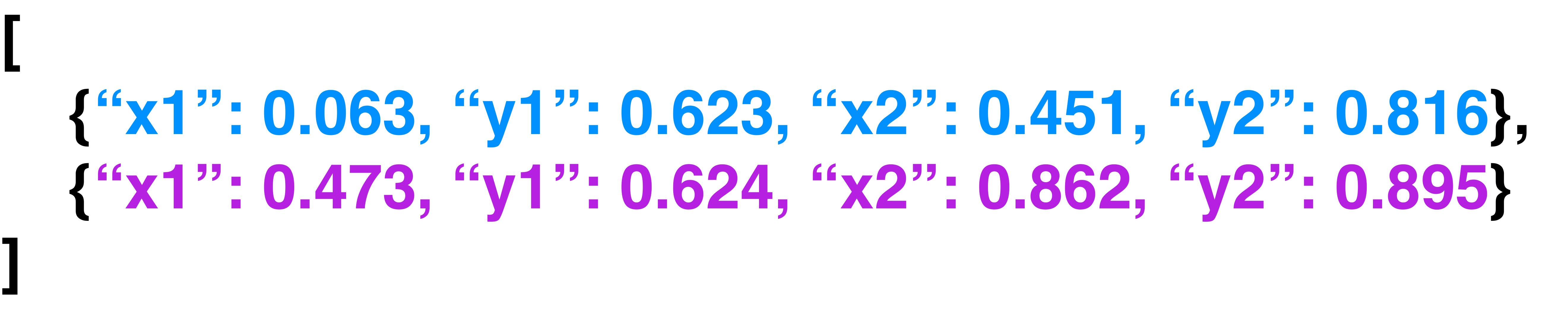}
        \caption{Exact Label}
        \label{fig:exact}
    \end{subfigure}
    \hfill
    \begin{subfigure}[h]{0.42\textwidth}
        \centering
        \includegraphics[width=\linewidth]{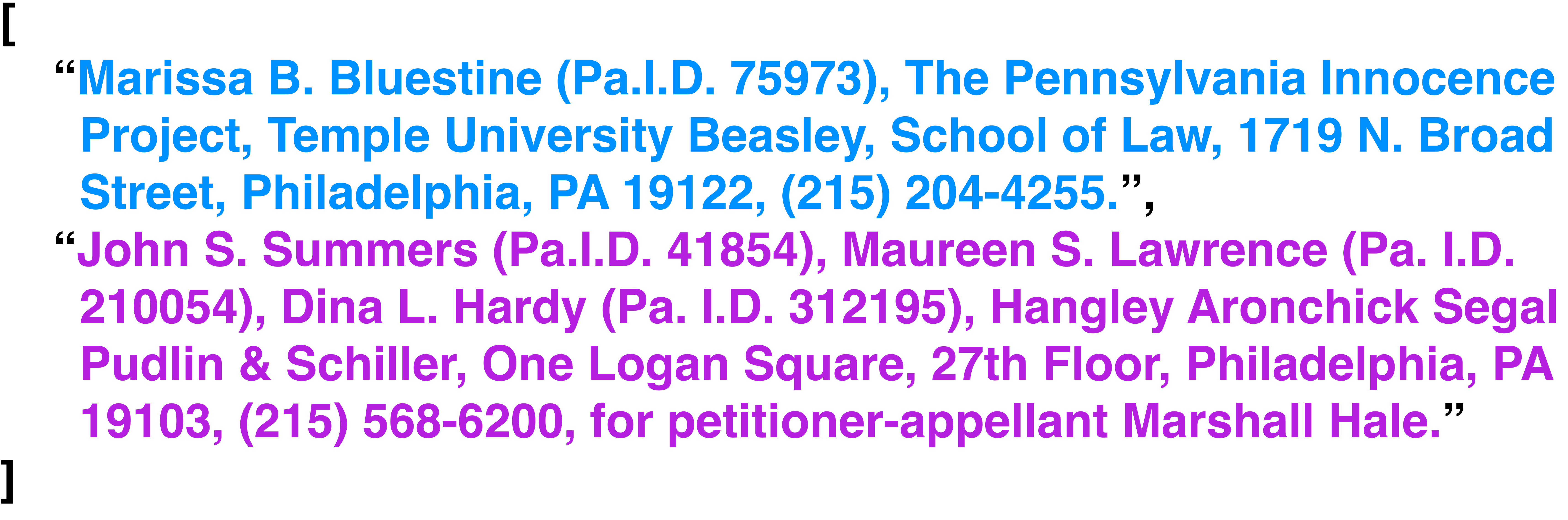}
        \caption{Inexact Label}
        \label{fig:inexact}
    \end{subfigure}
    \caption{An example of complex entity extraction from an appellate brief. (a) shows that the attorney profiles in the caption page violate the contiguity assumption for traditional NER problem, i.e., the OCR'd texts in (b) intertwine with each other. We address such an issue by a weakly supervised DLA approach, i.e., training an object detector to generate relevant bounding boxes as in (c), using image-level textual annotations as in (d).}
    \label{fig:example}
\end{figure}

In this work, we are focused on extracting the discontinuous complex NEs.  
Figure \ref{fig:brief} shows several NEs in the caption page of an appellate brief, 
including court name, parties' names, filed date, etc. 
Among these, the attorney profiles are an example of complex NEs, whose complication is introduced by Optical Character Recognition (OCR) \cite{lopresti2008optical}. 
Figure \ref{fig:ocr} depicts the problem: owing to the ambiguity in reading order, the OCR'd text on the left column (blue) intertwines with the text on the right column (purple), violating the contiguity assumption of the traditional NER methods, i.e., a NE mention should be a sequence of contiguous tokens \citep{dai2018recognizing}. 
To tackle this, we propose a novel system that incorporates an object detector for Document Layout Analysis (DLA) to alleviate the noise from OCR, especially in cases where the layout is complex. 

In addition, it should be emphasized that the solution is required to be efficient in terms of data usage.
Primarily, there are two practical reasons for this. First, the expense of annotating legal documents is significant, as it involves hiring legal professionals who typically charge a higher hourly rate. 
Second, frequent changes in business requirements necessitate constant modifications to labeling instructions, rendering recently annotated data imperfect.
In light of these concerns, we devise a weakly supervised method to leverage the available inexact labels.

To summarize, we combine DLA and heuristic rules to solve the complex NER problem. 
To the best of our knowledge, this is the first work applying weak supervision to DLA. 
The experimental results conclude that the system trained on pseudo labels can outperform the supervised baseline when gold-standard data is limited. 

\section{Related Work}
\subsection{NER in the legal domain} 
Legal NER demands tailored solutions due to the limited effectiveness of general-purpose NER methods in legal use cases, as noted by \citet{au2022ner}. 
Legal documents are characterized by longer sentence structures and lengthier NEs, presenting challenges to traditional NER methods.
Moreover, OCR errors pose a critical challenge in real-world scenarios, as a single misrecognized character can lead to misspellings (e.g., Claire or Clair) and out-of-vocabulary (OOV) words. To overcome these challenges, \citet{trias2021named} developed a name correction state machine based on heuristic rules, while \citet{skylaki2021legal} employed Pointer-Generator Networks to handle discontinuous entity mentions and OOV problems. Despite the complexity, capturing long and segmented named entities is a common business need, as seen in attorney profiles and the extraction of legal events \cite{filtz2020events}.

\subsection{Visually Rich Document Understanding} 
Business documents, such as invoices and legal documents, are a type of visually rich data that uses both textual information and visual features, such as layout and font style, to convey their message.
Thus, to effectively analyze such data, it requires utilizing a multi-modal approach that jointly learns the textual, visual, and layout knowledge in a single framework \cite{huang2022layoutlmv3, appalaraju2021docformer, powalski2021going}.
The related topics include Key Information Extraction (KIE) \cite{park2019cord, jaume2019funsd, huang2019icdar2019}, Document Layout Analysis (DLA) \cite{li2020docbank, zhong2019publaynet, pfitzmann2022doclaynet}, and DocVQA \cite{mathew2021docvqa}. 

In this work, we use DLA techniques, instead of the KIE ones that are mainly designed for the traditional NER problems (see more explanation in Section \ref{sec:def}). 
DLA is commonly conducted via object detection, which is well-suited for identifying the boundaries of the elements in a document.
It is independent of OCR, so it can be a remedy for the OCR errors. 
For our solution, we adopt LayoutLMv3 \cite{huang2022layoutlmv3}, which achieved state-of-the-art performance on PubLayNet \cite{zhong2019publaynet}, a popular DLA dataset. 
Compared with other DLA methods, the additional textual knowledge learned from pre-training advantages its predictions.

\subsection{Weakly Supervised Object Detection} 
Weakly supervised object detection techniques address the problem of inexact supervision \citep{zhang2021weakly, shao2022deep}. 
Instead of using object-level labels, these methods train an object detector using image-level labels, which only indicate the presence or absence of a specific object class in an image. In our work, we are only provided with the text of target NEs in a page, without corresponding bounding boxes, which is analogous to the aforementioned image-level labels. More information about our data is described in Section \ref{sec:pseudo-label}.

\section{Problem Definition}
\label{sec:def}
Traditionally, NER is formulated as a BIO sequence tagging problem that labels a sequence of tokens $x=\{w_i\}_{i=1}^N$ with a sequence of labels $y=\{q_i\}_{i=1}^{N}$, where $q_i \in \{B_t, I_t \mid t \in T\} \cup \{O\}$ and $B$, $I$, $O$ denote the \textbf{B}eginning, \textbf{I}nside, and \textbf{O}utside of a NE, respectively, for pre-defined NE types $T$. However, this tagging scheme assumes contiguous token occurrence for a NE, which does not hold in our scenario (see Figure~\ref{fig:ocr}).

Hence, we relax the above formulation and re-define the output $y$ as a set of unique token groups with predicted types, where each group is a subset of $x$ and each token is only used in one group: 
\begin{equation} 
    \label{eq:new-def}
    \begin{aligned}
        y &=\{G_e, t_e \mid e \in E\} \\
        G_e &= \{s_k \mid s \in x \}_{k=1}^{K_e} \textnormal{~and~} \forall u \ne v \in E, G_u \cap G_v = \varnothing\textnormal{;}
    \end{aligned}
\end{equation}
\noindent $t_e \in T$ denotes the type; $E$ is all the NEs; $K_e$ is the sequence length of $e$.
Note that token $s$ here is indexed by $k$ and the order can be different from their indices in $x$.  
In other words, the NEs are a set of unordered groups, whose tokens can be collected from arbitrary positions in a document, while the tokens within each group still must be in a specific order.

\section{Our Approach}
\begin{figure}[t!]
    \centering
    \includegraphics[width=\linewidth]{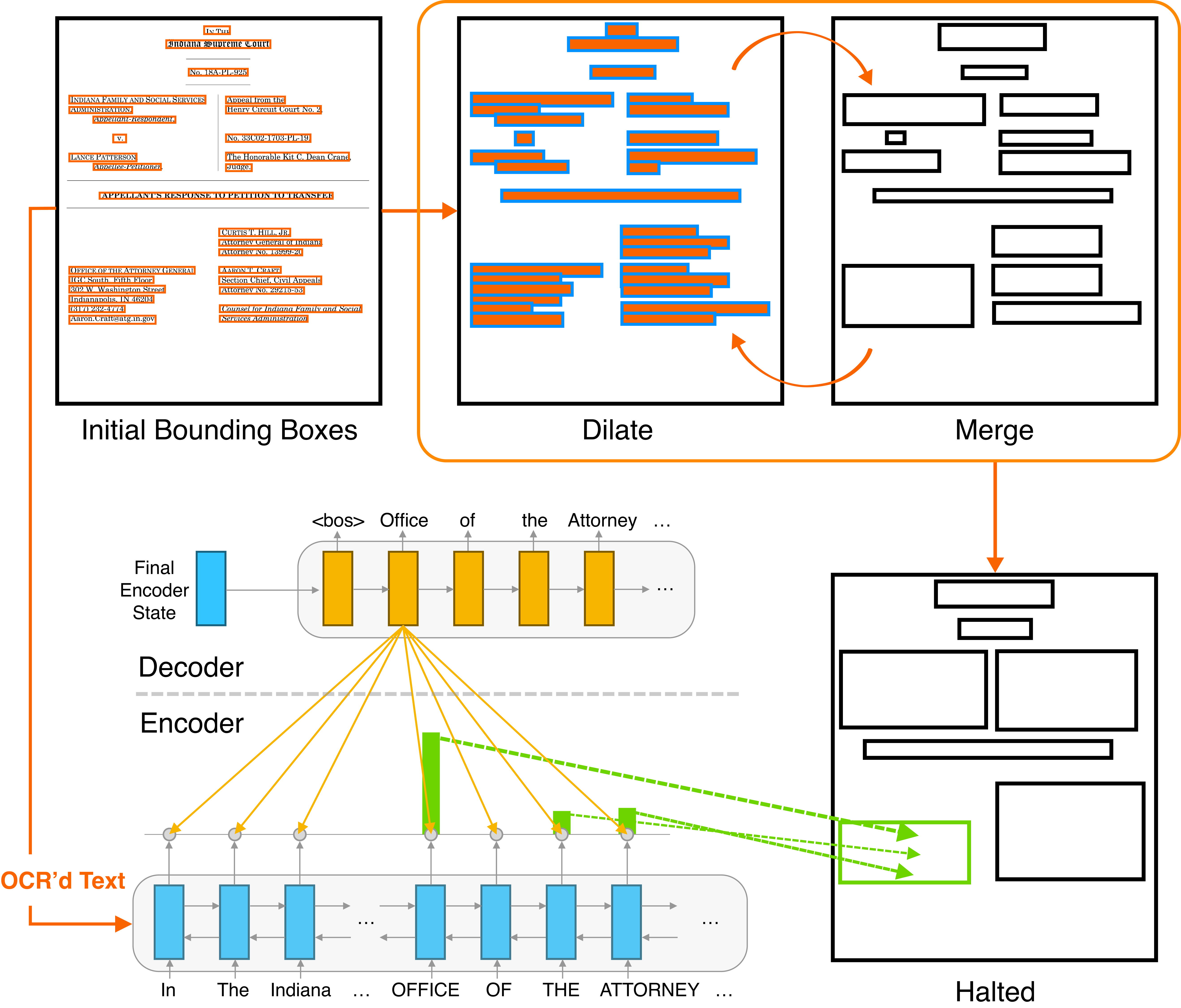}
    \caption{Object-level signal generation via paragraph segmentation and non-sequential token mapping.}
    \label{fig:system}
\end{figure}

To be more robust to complex layouts, we propose a multi-modal approach to extracting complex NEs in legal documents. 
The core of our solution is a weakly supervised object detector for DLA that directly captures relevant paragraphs from document images, before the OCR process. 
For training this component, we devise a pseudo label generator to create additional supervisory signal from textual inexact labels.
With the paragraphs being collected in a 2-D fashion and classified, we are allowed to established effective rules (see Section \ref{sec:post-processing}) to group and re-order the detected paragraphs; as a result, each group represents a NE in the final output.

\subsection{Exact and Inexact Labels}
\label{sec:pseudo-label}
For our object detection task, the exact labels are the bounding box (i.e., object-level) annotations of the target NEs on document images, from which we can calculate the overlapped area with the word-level bounding boxes from OCR to map the text. 
Figure~\ref{fig:exact} is an example of the exact labels for attorney profile. 
By contrast, the inexact labels are the image-level texts of the NEs, without location information (i.e., the corresponding bounding boxes).
Their literal values are not necessarily copied from the documents, owing to the previous business requirements of reformation.
For instance, as shown in Figure~\ref{fig:inexact}, the annotators were asked to compile the sub-elements of attorney profile in the following order: attorney name, title, bar number, firm name, address, city, state, phone number, fax number, e-mail address, and party designation.
Moreover, the case of words was changed to title case, and extra commas were placed manually to separate the sub-elements. 

\subsection{Pseudo Label Generator}
To produce more object-level signals, we developed a 2-stage pseudo label generator based on our inexact labels. 
In the initial stage, we create Regions of Interest (RoI) proposals by applying paragraph segmentation to a document. 
In the subsequent stage, a RoI selector that is trained on our inexact labels is used to filter the proposals. 
More specifically, this selector learns a non-sequential token mapping between the inexact labels and the OCR'd text, 
by which we assess how relevant a proposal is to the target NEs. 
Finally, our object detector is fed the selected RoIs to perform weakly supervised learning.
We explain more details as follows: 

\subsubsection{Paragraph Segmentation}
Although the NE mentions in 1-D OCR'd texts can be interrupted by noise, they are still mostly laid out as paragraphs in document images, from a 2-D perspective. 
According to such observation, we devise a morphology-based \cite{doermann2014handbook} algorithm to perform paragraph segmentation.
The input is a set of initial bounding boxes $B_0$, and the algorithm yields a set of paragraph RoIs $B_h$ after $h$ times of iteration.
The proposed technique is depicted in the upper part of Figure \ref{fig:system}.
To summarize, it dilates the bounding boxes with a kernel $\mathcal{K}$ and merges the overlapped ones repeatedly. After each iteration, the RoI of a paragraph is re-defined by the contour of the merged lines. This process continues until one of the two halting conditions is reached.
The first one is patience $\rho$: the process is halted if no merger is triggered after $\rho$ iterations.
The other is aggregation ratio $\delta=\frac{|B_h|}{|B_0|}$, which restricts the minimum number of RoIs.

\subsubsection{Non-sequential Token Mapping}
To enable RoI selection, we need to measure the relevance scores of RoIs based on token coverage,  
i.e., a RoI that has more tokens that are used in target NEs should receive a higher score. 
However, our inexact labels only provides the literal values of NEs. 
Without knowing the 2-D locations, we are not able to directly calculate this coverage.

To bridge this gap, we train a sequence-to-sequence model to infer a mapping from inexact labels to OCR'd text, from which we can obtain the associated bounding boxes. 
Note that inexact labels and OCR'd text can have inconsistent token ordering (e.g., when inexact labels are re-ordered as in Figure \ref{fig:inexact}), so tracing the locations via a sequence matching algorithm might not be always feasible.
Inspired by~\citet{skylaki2021legal}, who exhibited the effectiveness of Pointer-Generator Networks \cite{see2017get} on legal NER, we exploit the same model and use its copying mechanism as the non-sequential token mapping.
The source (target) sequence for this model is the OCR'd text (the inexact labels of the target NEs). 
We observed that the trained model behaves similarly to human annotators --- the majority of output tokens are copied from the document; 
thus, we can rely on the attention score $A_{i,j}$ to estimate the location, where $i$ ($j$) denotes the encoder (decoder) step.  
The lower part of Figure \ref{fig:system} illustrates the model and how the attention scores can aid the coverage estimation. 

\subsubsection{Object-Level Signal Generation.}
By using the produced RoIs (i.e., paragraph bounding boxes) $B_h$ and the token mapping $A_{i,j}$, we can measure the relevancy of the proposals and preserve the ones satisfying the pre-defined thresholds. 
In order to be compatible with our internal explainability tool, we first aggregate the attention scores to line level $A_{l,j} = \sum_{i=l_s}^{l_e}A_{i,j}$, where $l$ is a line and $l_s$ ($l_e$) denotes the starting (ending) index. A line $l$ is considered active if $A_{l,j}/N_l > \psi$, and we preserve the paragraphs which have more than $\phi$ active lines, where $N_l$ is the number of words in $l$; $\psi$ and $\phi$ are the thresholds.

\section{Experiments}
In this paper, we test our approach on extracting attorney profiles from appellate briefs. This is one of the most time-consuming processes in our manual workflow, because the profile information is more detailed in this document type, and the amount can exceed 170K documents per year. 
At the same time, it is also a challenging machine learning problem due to being complexly laid out, having potential to drive future innovation.
In the following, we explain our experiments further: 

\subsection{Experimental Settings}
\subsubsection{Datasets.} 
Our dataset consists of approximately 600K legal briefs (in PDF format) across 166 courts in the U.S., dating from 2014 to 2020. 
The texts are OCR'd by AWS Textract,\footnote{\url{https://aws.amazon.com/textract/}} provided with word- and line- level bounding boxes. We have 3 sets of labels: 
\begin{itemize}
    \item \textbf{Inexact}: all the briefs are attached with the inexact labels, as shown in Figure~\ref{fig:inexact};
    \item \textbf{Gold}: 709 documents are annotated as in Figure~\ref{fig:exact} (i.e., exact label) by a team of 5 subject matter experts. The group index and token order are specified, allowing us to unambiguously derive NEs that meet Definition~\ref{eq:new-def}. These labels are split into training/validation/test sets at 40\%/10\%/50\%, respectively;
    \item \textbf{Pseudo}: 4k documents (i.e., 15 times larger than the training set of the exact labels) are separately sampled to be inferred by the proposed approach. 
\end{itemize}

\subsubsection{Model Configurations}
We trained the Pointer-Generator Networks with inexact labels. Notable configuration changes are input/output lengths, which are set to 3,000/400, respectively.
Our object detector, LayoutLMv3 combined with Casade R-CNN \cite{cai2018cascade}, was fine-tuned in three different settings for comparative analysis. They are fine-tuned with Gold, Pseudo, and Pseudo+Gold labels from the checkpoint that was fine-tuned on PubLayNet. The paragraph- (line-) threshold $\phi$ ($\psi$) is 1 (0.1). See Appendix \ref{appx:model-config} for the rest.

\subsubsection{Post-processing}
\label{sec:post-processing}
To provide final NE values that meet Definition~\ref{eq:new-def}, we applied rule-based processes of grouping and re-ordering. 
Since the objects are already classified, we are allowed to build effective rules based on stronger assumptions. 
For attorney profile, we impose column-major order\footnote{\url{https://en.wikipedia.org/wiki/Row-_and_column-major_order}} by implementing X-cut \cite{ha1995recursive}, and grouping is determined by their sizes and designation lines, e.g., “Attorneys for Petitioner”. See Appendix \ref{appx:post-processing} for more details.

\subsection{Results and Discussion}
We adopted \emph{exact match} to evaluate our system outputs and present the results in Table~\ref{tab:scores}.
Note that the three settings were evaluated on Gold-test labels and required no post-processing. 
By comparison, it shows that the weakly supervised object detector trained on Pseudo labels outperformed the supervised model trained on Gold labels. 
This indicates that our method can effectively produce high-quality signals, and Pseudo labels are comparably informative as Gold labels. 
Furthermore, we conducted an ablation analysis by comparing the models trained on Pseudo and Pseudo+Gold labels.
Surprisingly, additional provision of Gold labels does not improve our solution pipeline. 
This finding suggests that Pseudo labels can sufficiently cover the information embedded in Gold labels. 
We additionally report average \emph{Edit Distance} to reflect the expenses incurred when the systems are deployed to a human-in-the-loop workflow. 
The results are similar in all three settings and require approximately five words to be corrected.

\begin{figure}[t!]
    \centering
    \includegraphics[width=\linewidth]{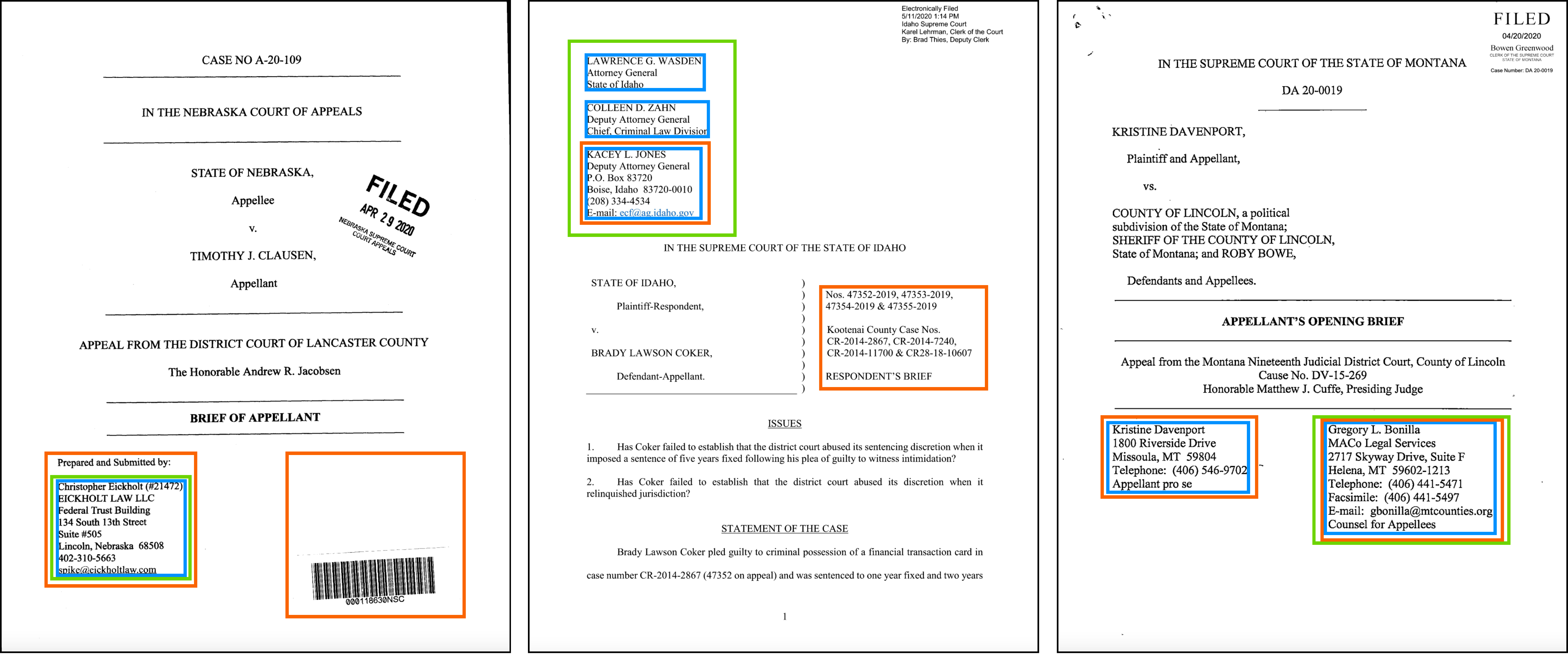}
    \caption{Case Study. Comparison among ground-truth (green), Gold model (orange), and Pseudo model (blue).}
    \label{fig:case}
\end{figure}

Figure \ref{fig:case} presents a case study of errors made by the Gold model. On the left, it tends to produce larger objects and incorrectly places one on the barcode, possibly due to a bias caused by limited data and over-sensitivity to common locations, such as the bottom-right area. In the middle, it misses the upper objects, and we attribute to a similar reason -- fewer cases have attorney profiles at the top. On the right, both the Gold and Pseudo models produce false positives for the \emph{pro se} profile. This is a tricky problem, since it is excluded by business requirements. The inability of the object detector to read text prevents it from recognizing this pattern. Nonetheless, a simple text-based rule should be able to compensate for this.

\begin{table}[t!]\small
  \caption{Results of Attorney Profile Extraction: P/R/F1 stand for Precision/Recall/F1-score, respectively.}
  \label{tab:scores}
  \begin{tabular}{l|cccc}
    \toprule
    \textbf{Dataset} & \textbf{P} & \textbf{R} & \textbf{F1} & \textbf{Edit Distance} \\
    \midrule
    Gold & 0.637 & 0.633 & 0.635 & 5.181 \\
    Pseudo & \textbf{0.665} & \textbf{0.643} & \textbf{0.654} & \textbf{5.135} \\
    Pseudo+Gold & \textbf{0.665} & 0.635 & 0.650 & 5.711 \\
    \bottomrule
  \end{tabular}
\end{table}

\section{Conclusion}
In this work, we have demonstrated an effective weakly supervised approach to extracting complex NEs from legal documents.
Considering the substantial amount of data to process, our solution has the potential to significantly reduce costs and save time.

\section*{Company Portrait}
\textbf{Thomson Reuters} is best known for Reuters News but also the leading information provider for legal, corporate, tax, and accounting professionals. They have 60,000 TBs of data and Thomson Reuters Labs has been innovative in AI for 30 years.

\section*{Presenter}
\textbf{Hsiu-Wei Yang} is an Applied Research Scientist at Thomson Reuters Labs in Toronto, Canada. He has a Master's degree in Computer Science from University of Waterloo. His research focuses on Information Retrieval and Natural Language Processing.

\newpage

\bibliographystyle{ACM-Reference-Format}
\bibliography{main}


\begin{thebibliography}{26}

%%% the Web address.

\ifx \showCODEN    \undefined \def \showCODEN     #1{\unskip}     \fi
\ifx \showDOI      \undefined \def \showDOI       #1{#1}\fi
\ifx \showISBNx    \undefined \def \showISBNx     #1{\unskip}     \fi
\ifx \showISBNxiii \undefined \def \showISBNxiii  #1{\unskip}     \fi
\ifx \showISSN     \undefined \def \showISSN      #1{\unskip}     \fi
\ifx \showLCCN     \undefined \def \showLCCN      #1{\unskip}     \fi
\ifx \shownote     \undefined \def \shownote      #1{#1}          \fi
\ifx \showarticletitle \undefined \def \showarticletitle #1{#1}   \fi
\ifx \showURL      \undefined \def \showURL       {\relax}        \fi
% invisible to TeX
\providecommand\bibfield[2]{#2}
\providecommand\bibinfo[2]{#2}
\providecommand\natexlab[1]{#1}
\providecommand\showeprint[2][]{arXiv:#2}

\bibitem[Appalaraju et~al\mbox{.}(2021)]%
        {appalaraju2021docformer}
\bibfield{author}{\bibinfo{person}{Srikar Appalaraju}, \bibinfo{person}{Bhavan
  Jasani}, \bibinfo{person}{Bhargava~Urala Kota}, \bibinfo{person}{Yusheng
  Xie}, {and} \bibinfo{person}{R. Manmatha}.} \bibinfo{year}{2021}\natexlab{}.
\newblock \showarticletitle{DocFormer: End-to-End Transformer for Document
  Understanding}. In \bibinfo{booktitle}{\emph{Proceedings of the IEEE/CVF
  International Conference on Computer Vision (ICCV)}}.
  \bibinfo{pages}{993--1003}.
\newblock


\bibitem[Au et~al\mbox{.}(2022)]%
        {au2022ner}
\bibfield{author}{\bibinfo{person}{Ting Wai~Terence Au},
  \bibinfo{person}{Vasileios Lampos}, {and} \bibinfo{person}{Ingemar Cox}.}
  \bibinfo{year}{2022}\natexlab{}.
\newblock \showarticletitle{{E}-{NER} {---} An Annotated Named Entity
  Recognition Corpus of Legal Text}. In \bibinfo{booktitle}{\emph{Proceedings
  of the Natural Legal Language Processing Workshop 2022}}.
  \bibinfo{publisher}{Association for Computational Linguistics},
  \bibinfo{pages}{246--255}.
\newblock


\bibitem[Cai and Vasconcelos(2018)]%
        {cai2018cascade}
\bibfield{author}{\bibinfo{person}{Zhaowei Cai} {and} \bibinfo{person}{Nuno
  Vasconcelos}.} \bibinfo{year}{2018}\natexlab{}.
\newblock \showarticletitle{Cascade r-cnn: Delving into high quality object
  detection}. In \bibinfo{booktitle}{\emph{Proceedings of the IEEE conference
  on computer vision and pattern recognition}}. \bibinfo{pages}{6154--6162}.
\newblock


\bibitem[Dai(2018)]%
        {dai2018recognizing}
\bibfield{author}{\bibinfo{person}{Xiang Dai}.}
  \bibinfo{year}{2018}\natexlab{}.
\newblock \showarticletitle{Recognizing complex entity mentions: A review and
  future directions}. In \bibinfo{booktitle}{\emph{Proceedings of ACL 2018,
  Student Research Workshop}}. \bibinfo{pages}{37--44}.
\newblock


\bibitem[Doermann et~al\mbox{.}(2014)]%
        {doermann2014handbook}
\bibfield{author}{\bibinfo{person}{David Doermann}, \bibinfo{person}{Karl
  Tombre}, {et~al\mbox{.}}} \bibinfo{year}{2014}\natexlab{}.
\newblock \bibinfo{booktitle}{\emph{Handbook of document image processing and
  recognition}}. Vol.~\bibinfo{volume}{1}.
\newblock \bibinfo{publisher}{Springer}.
\newblock


\bibitem[Filtz et~al\mbox{.}(2020)]%
        {filtz2020events}
\bibfield{author}{\bibinfo{person}{Erwin Filtz}, \bibinfo{person}{Mar{\'\i}a
  Navas-Loro}, \bibinfo{person}{Cristiana Santos}, \bibinfo{person}{Axel
  Polleres}, {and} \bibinfo{person}{Sabrina Kirrane}.}
  \bibinfo{year}{2020}\natexlab{}.
\newblock \showarticletitle{Events matter: Extraction of events from court
  decisions}.
\newblock \bibinfo{journal}{\emph{Legal Knowledge and Information Systems}}
  (\bibinfo{year}{2020}), \bibinfo{pages}{33--42}.
\newblock


\bibitem[Guo et~al\mbox{.}(2009)]%
        {guo2009named}
\bibfield{author}{\bibinfo{person}{Jiafeng Guo}, \bibinfo{person}{Gu Xu},
  \bibinfo{person}{Xueqi Cheng}, {and} \bibinfo{person}{Hang Li}.}
  \bibinfo{year}{2009}\natexlab{}.
\newblock \showarticletitle{Named entity recognition in query}. In
  \bibinfo{booktitle}{\emph{Proceedings of the 32nd international ACM SIGIR
  conference on Research and development in information retrieval}}.
  \bibinfo{pages}{267--274}.
\newblock


\bibitem[Ha et~al\mbox{.}(1995)]%
        {ha1995recursive}
\bibfield{author}{\bibinfo{person}{Jaekyu Ha}, \bibinfo{person}{Robert~M
  Haralick}, {and} \bibinfo{person}{Ihsin~T Phillips}.}
  \bibinfo{year}{1995}\natexlab{}.
\newblock \showarticletitle{Recursive XY cut using bounding boxes of connected
  components}. In \bibinfo{booktitle}{\emph{Proceedings of 3rd International
  Conference on Document Analysis and Recognition}}, Vol.~\bibinfo{volume}{2}.
  IEEE, \bibinfo{pages}{952--955}.
\newblock


\bibitem[Huang et~al\mbox{.}(2022)]%
        {huang2022layoutlmv3}
\bibfield{author}{\bibinfo{person}{Yupan Huang}, \bibinfo{person}{Tengchao Lv},
  \bibinfo{person}{Lei Cui}, \bibinfo{person}{Yutong Lu}, {and}
  \bibinfo{person}{Furu Wei}.} \bibinfo{year}{2022}\natexlab{}.
\newblock \showarticletitle{Layoutlmv3: Pre-training for document ai with
  unified text and image masking}. In \bibinfo{booktitle}{\emph{Proceedings of
  the 30th ACM International Conference on Multimedia}}.
  \bibinfo{pages}{4083--4091}.
\newblock


\bibitem[Huang et~al\mbox{.}(2019)]%
        {huang2019icdar2019}
\bibfield{author}{\bibinfo{person}{Zheng Huang}, \bibinfo{person}{Kai Chen},
  \bibinfo{person}{Jianhua He}, \bibinfo{person}{Xiang Bai},
  \bibinfo{person}{Dimosthenis Karatzas}, \bibinfo{person}{Shijian Lu}, {and}
  \bibinfo{person}{CV Jawahar}.} \bibinfo{year}{2019}\natexlab{}.
\newblock \showarticletitle{Icdar2019 competition on scanned receipt ocr and
  information extraction}. In \bibinfo{booktitle}{\emph{2019 International
  Conference on Document Analysis and Recognition (ICDAR)}}. IEEE,
  \bibinfo{pages}{1516--1520}.
\newblock


\bibitem[Jaume et~al\mbox{.}(2019)]%
        {jaume2019funsd}
\bibfield{author}{\bibinfo{person}{Guillaume Jaume},
  \bibinfo{person}{Hazim~Kemal Ekenel}, {and} \bibinfo{person}{Jean-Philippe
  Thiran}.} \bibinfo{year}{2019}\natexlab{}.
\newblock \showarticletitle{Funsd: A dataset for form understanding in noisy
  scanned documents}. In \bibinfo{booktitle}{\emph{2019 International
  Conference on Document Analysis and Recognition Workshops (ICDARW)}},
  Vol.~\bibinfo{volume}{2}. IEEE, \bibinfo{pages}{1--6}.
\newblock


\bibitem[Khalid et~al\mbox{.}(2008)]%
        {khalid2008impact}
\bibfield{author}{\bibinfo{person}{Mahboob~Alam Khalid},
  \bibinfo{person}{Valentin Jijkoun}, {and} \bibinfo{person}{Maarten
  De~Rijke}.} \bibinfo{year}{2008}\natexlab{}.
\newblock \showarticletitle{The impact of named entity normalization on
  information retrieval for question answering}. In
  \bibinfo{booktitle}{\emph{Advances in Information Retrieval: 30th European
  Conference on IR Research, ECIR 2008, Glasgow, UK, March 30-April 3, 2008.
  Proceedings 30}}. Springer, \bibinfo{pages}{705--710}.
\newblock


\bibitem[Li et~al\mbox{.}(2020)]%
        {li2020docbank}
\bibfield{author}{\bibinfo{person}{Minghao Li}, \bibinfo{person}{Yiheng Xu},
  \bibinfo{person}{Lei Cui}, \bibinfo{person}{Shaohan Huang},
  \bibinfo{person}{Furu Wei}, \bibinfo{person}{Zhoujun Li}, {and}
  \bibinfo{person}{Ming Zhou}.} \bibinfo{year}{2020}\natexlab{}.
\newblock \showarticletitle{{D}oc{B}ank: A Benchmark Dataset for Document
  Layout Analysis}. In \bibinfo{booktitle}{\emph{Proceedings of the 28th
  International Conference on Computational Linguistics}}.
  \bibinfo{pages}{949--960}.
\newblock


\bibitem[Lopresti(2008)]%
        {lopresti2008optical}
\bibfield{author}{\bibinfo{person}{Daniel Lopresti}.}
  \bibinfo{year}{2008}\natexlab{}.
\newblock \showarticletitle{Optical character recognition errors and their
  effects on natural language processing}. In
  \bibinfo{booktitle}{\emph{Proceedings of the second workshop on Analytics for
  Noisy Unstructured Text Data}}. \bibinfo{pages}{9--16}.
\newblock


\bibitem[Malmasi et~al\mbox{.}(2022)]%
        {malmasi2022semeval}
\bibfield{author}{\bibinfo{person}{Shervin Malmasi}, \bibinfo{person}{Anjie
  Fang}, \bibinfo{person}{Besnik Fetahu}, \bibinfo{person}{Sudipta Kar}, {and}
  \bibinfo{person}{Oleg Rokhlenko}.} \bibinfo{year}{2022}\natexlab{}.
\newblock \showarticletitle{Semeval-2022 task 11: Multilingual complex named
  entity recognition (multiconer)}. In \bibinfo{booktitle}{\emph{Proceedings of
  the 16th international workshop on semantic evaluation (SemEval-2022)}}.
  \bibinfo{pages}{1412--1437}.
\newblock


\bibitem[Mathew et~al\mbox{.}(2021)]%
        {mathew2021docvqa}
\bibfield{author}{\bibinfo{person}{Minesh Mathew}, \bibinfo{person}{Dimosthenis
  Karatzas}, {and} \bibinfo{person}{CV Jawahar}.}
  \bibinfo{year}{2021}\natexlab{}.
\newblock \showarticletitle{Docvqa: A dataset for vqa on document images}. In
  \bibinfo{booktitle}{\emph{Proceedings of the IEEE/CVF winter conference on
  applications of computer vision}}. \bibinfo{pages}{2200--2209}.
\newblock


\bibitem[Park et~al\mbox{.}(2019)]%
        {park2019cord}
\bibfield{author}{\bibinfo{person}{Seunghyun Park}, \bibinfo{person}{Seung
  Shin}, \bibinfo{person}{Bado Lee}, \bibinfo{person}{Junyeop Lee},
  \bibinfo{person}{Jaeheung Surh}, \bibinfo{person}{Minjoon Seo}, {and}
  \bibinfo{person}{Hwalsuk Lee}.} \bibinfo{year}{2019}\natexlab{}.
\newblock \showarticletitle{CORD: A Consolidated Receipt Dataset for Post-OCR
  Parsing}.
\newblock  (\bibinfo{year}{2019}).
\newblock


\bibitem[Pfitzmann et~al\mbox{.}(2022)]%
        {pfitzmann2022doclaynet}
\bibfield{author}{\bibinfo{person}{Birgit Pfitzmann},
  \bibinfo{person}{Christoph Auer}, \bibinfo{person}{Michele Dolfi},
  \bibinfo{person}{Ahmed~S Nassar}, {and} \bibinfo{person}{Peter Staar}.}
  \bibinfo{year}{2022}\natexlab{}.
\newblock \showarticletitle{DocLayNet: A Large Human-Annotated Dataset for
  Document-Layout Segmentation}. In \bibinfo{booktitle}{\emph{Proceedings of
  the 28th ACM SIGKDD Conference on Knowledge Discovery and Data Mining}}.
  \bibinfo{pages}{3743--3751}.
\newblock


\bibitem[Powalski et~al\mbox{.}(2021)]%
        {powalski2021going}
\bibfield{author}{\bibinfo{person}{Rafa{\l} Powalski},
  \bibinfo{person}{{\L}ukasz Borchmann}, \bibinfo{person}{Dawid Jurkiewicz},
  \bibinfo{person}{Tomasz Dwojak}, \bibinfo{person}{Micha{\l} Pietruszka},
  {and} \bibinfo{person}{Gabriela Pa{\l}ka}.} \bibinfo{year}{2021}\natexlab{}.
\newblock \showarticletitle{Going full-tilt boogie on document understanding
  with text-image-layout transformer}. In \bibinfo{booktitle}{\emph{Document
  Analysis and Recognition--ICDAR 2021: 16th International Conference,
  Lausanne, Switzerland, September 5--10, 2021, Proceedings, Part II 16}}.
  Springer, \bibinfo{pages}{732--747}.
\newblock


\bibitem[See et~al\mbox{.}(2017)]%
        {see2017get}
\bibfield{author}{\bibinfo{person}{Abigail See}, \bibinfo{person}{Peter~J.
  Liu}, {and} \bibinfo{person}{Christopher~D. Manning}.}
  \bibinfo{year}{2017}\natexlab{}.
\newblock \showarticletitle{Get To The Point: Summarization with
  Pointer-Generator Networks}. In \bibinfo{booktitle}{\emph{Proceedings of the
  55th Annual Meeting of the Association for Computational Linguistics (Volume
  1: Long Papers)}}. \bibinfo{pages}{1073--1083}.
\newblock


\bibitem[Shao et~al\mbox{.}(2022)]%
        {shao2022deep}
\bibfield{author}{\bibinfo{person}{Feifei Shao}, \bibinfo{person}{Long Chen},
  \bibinfo{person}{Jian Shao}, \bibinfo{person}{Wei Ji},
  \bibinfo{person}{Shaoning Xiao}, \bibinfo{person}{Lu Ye},
  \bibinfo{person}{Yueting Zhuang}, {and} \bibinfo{person}{Jun Xiao}.}
  \bibinfo{year}{2022}\natexlab{}.
\newblock \showarticletitle{Deep learning for weakly-supervised object
  detection and localization: A survey}.
\newblock \bibinfo{journal}{\emph{Neurocomputing}}  \bibinfo{volume}{496}
  (\bibinfo{year}{2022}), \bibinfo{pages}{192--207}.
\newblock


\bibitem[Skylaki et~al\mbox{.}(2021)]%
        {skylaki2021legal}
\bibfield{author}{\bibinfo{person}{Stavroula Skylaki}, \bibinfo{person}{Ali
  Oskooei}, \bibinfo{person}{Omar Bari}, \bibinfo{person}{Nadja Herger}, {and}
  \bibinfo{person}{Zac Kriegman}.} \bibinfo{year}{2021}\natexlab{}.
\newblock \showarticletitle{Legal Entity Extraction using a Pointer Generator
  Network}. In \bibinfo{booktitle}{\emph{2021 International Conference on Data
  Mining Workshops (ICDMW)}}. IEEE, \bibinfo{pages}{653--658}.
\newblock


\bibitem[Trias et~al\mbox{.}(2021)]%
        {trias2021named}
\bibfield{author}{\bibinfo{person}{Fernando Trias}, \bibinfo{person}{Hongming
  Wang}, \bibinfo{person}{Sylvain Jaume}, {and} \bibinfo{person}{Stratos
  Idreos}.} \bibinfo{year}{2021}\natexlab{}.
\newblock \showarticletitle{Named entity recognition in historic legal text: A
  transformer and state machine ensemble method}. In
  \bibinfo{booktitle}{\emph{Proceedings of the Natural Legal Language
  Processing Workshop 2021}}. \bibinfo{pages}{172--179}.
\newblock


\bibitem[Voorhees et~al\mbox{.}(2005)]%
        {voorhees2005trec}
\bibfield{author}{\bibinfo{person}{Ellen~M Voorhees}, \bibinfo{person}{Donna~K
  Harman}, {et~al\mbox{.}}} \bibinfo{year}{2005}\natexlab{}.
\newblock \bibinfo{booktitle}{\emph{TREC: Experiment and evaluation in
  information retrieval}}. Vol.~\bibinfo{volume}{63}.
\newblock \bibinfo{publisher}{Citeseer}.
\newblock


\bibitem[Zhang et~al\mbox{.}(2021)]%
        {zhang2021weakly}
\bibfield{author}{\bibinfo{person}{Dingwen Zhang}, \bibinfo{person}{Junwei
  Han}, \bibinfo{person}{Gong Cheng}, {and} \bibinfo{person}{Ming-Hsuan Yang}.}
  \bibinfo{year}{2021}\natexlab{}.
\newblock \showarticletitle{Weakly supervised object localization and
  detection: A survey}.
\newblock \bibinfo{journal}{\emph{IEEE transactions on pattern analysis and
  machine intelligence}} \bibinfo{volume}{44}, \bibinfo{number}{9}
  (\bibinfo{year}{2021}), \bibinfo{pages}{5866--5885}.
\newblock


\bibitem[Zhong et~al\mbox{.}(2019)]%
        {zhong2019publaynet}
\bibfield{author}{\bibinfo{person}{Xu Zhong}, \bibinfo{person}{Jianbin Tang},
  {and} \bibinfo{person}{Antonio~Jimeno Yepes}.}
  \bibinfo{year}{2019}\natexlab{}.
\newblock \showarticletitle{Publaynet: largest dataset ever for document layout
  analysis}. In \bibinfo{booktitle}{\emph{2019 International Conference on
  Document Analysis and Recognition (ICDAR)}}. IEEE,
  \bibinfo{pages}{1015--1022}.
\newblock


\end{thebibliography}

\appendix
\section{Details of Model Configurations}
\label{appx:model-config}
\begin{itemize}
    \item \textbf{Non-sequential Token Mapping}: we trained the Pointer-Generator Networks, implemented by OpenNMT\footnote{\url{https://opennmt.net/OpenNMT-py/}}, with inexact labels. Input and output lengths are changed to 3000 and 400, respectively; batch size is 2; the rest hyper-parameters are set as default. All the documents except for the ones that have gold labels are used, out of which 10K are kept for validation. Totally, it is trained by 780k steps.
    \item \textbf{Object Detector}: following \citet{huang2022layoutlmv3}, LayoutLMv3 feature backbone and Cascade R-CNN \cite{cai2018cascade} are integrated to perform object detection. We use the same configuration and start from the checkpoint fine-tuned on PubLayNet \cite{zhong2019publaynet}. To compare the performance, we further fine-tuned it on gold, pseudo, and pseudo+gold labels, individually. 
    Confidence threshold is set to 0.5. When there are overlapped objects, only the largest one is returned.
    \item \textbf{Others}: kernel size $\mathcal{K}$ is (2,2), patience $\rho$ is 3, aggregation rate $\delta$ is 0.5, and paragraph- (line-) threshold $\phi$ ($\psi$) is 1 (0.1).
\end{itemize}

\section{Details of Post-processing}
\label{appx:post-processing}
%TODO: say why it's more reasonable
%TODO: add some exceptions 
To obtain the NE results that meet Definition \ref{eq:new-def}, the last process is to separate the collected tokens covered by the predicted objects into groups and determine the token order within each group. 
We devised a rule-based and type-specific process to achieve this.

For attorney profile, we first analyzed and added three attributes to each predicted object: (1) \emph{isMajor}: if an object has more than five lines; (2) \emph{isDesignation}: if the text of an object has the word \emph{for}, e.g., ``Attorneys \emph{for} Petitioner''; (3) \emph{columnIndex}: the index of the column to which an object belongs, decided by X-cut, i.e., similar to XY-cut \cite{ha1995recursive} but we only analyze x-axis. 
With these attributes along with their coordinates, we sort them in column-major order, which aligns with how attorney profiles are usually read. 
Furthermore, within each column, we merge minor objects (i.e., \emph{isMajor=False}) into the nearest major ones, but however when a major object is or becomes a designation object (i.e., \emph{isDesignation=True}) it stops accepting new minor members. 
As a result, each object group represents a NE as our final output.

\end{document}